\documentclass[twocolumn,preprintnumbers,amsmath,amssymb]{revtex4}
\usepackage{graphicx}
\usepackage{dcolumn}
\usepackage{bm}

\usepackage{xcolor}
\usepackage{multirow}

\newcommand{\be}{\begin{equation}}
\newcommand{\ee}{\end{equation}}
\newcommand{\bea}{\begin{eqnarray}}
\newcommand{\eea}{\end{eqnarray}}

\newcommand{\tw}{\rm w}

\newcommand{\cN}{{\mathcal N}}

\begin{document}

\preprint{CERN-TH-2019-082}

\title{Swampland Bounds on the Abelian Gauge Sector}
\author{Seung-Joo Lee${}^{1}$ and Timo Weigand${}^{1,2}$}
\email[]{seung.joo.lee@cern.ch, timo.weigand @cern.ch}
\affiliation{%
\small ${}^{1}$CERN Theory Department, 1 Esplanade des Particules, CH-1211 Geneva, Switzerland \\
\small ${}^{2}${\it PRISMA Cluster of Excellence and Mainz Institute for Theoretical Physics, \\
Johannes Gutenberg-Universit\"at, 55099 Mainz, Germany}}


\begin{abstract}

We derive bounds on the number of abelian gauge group factors in six-dimensional gravitational theories with minimal supersymmetry and in their F-theoretic realisations.
These bounds follow by requiring consistency of certain BPS strings in the spectrum of the theory, as recently proposed in the literature.
Under certain assumptions this approach constrains the number of abelian gauge group factors in six-dimensional supergravity theories with at least one tensor multiplet to be $N \leq 20$ (or $N \leq 22$ in absence of charged matter).
For any geometric F-theory realisation with at least one tensor multiplet we establish the bound $N \leq 16$ by demanding unitarity of a heterotic solitonic string which exists even in absence of a perturbative heterotic dual.
This result extends to four-dimensional F-theory vacua on any blowup of a rational fibration. Our findings lead to universal bounds on the rank of the Mordell-Weil group of elliptically fibered Calabi-Yau threefolds.

\end{abstract}
\maketitle

\section{Introduction}

It is widely believed that compatibility of a gauge theory with quantum gravity imposes extra constraints on the structure of the theory which may not be 
visible from consistency of the low-energy effective theory alone. The resulting schism of theory space into a swampland \cite{Vafa:2005ui} of theories lacking an embedding into quantum gravity versus the landscape 
of fully consistent theories including gravity is a subject that has attracted considerable recent attention \cite{Brennan:2017rbf,Palti:2019pca}. 

In this note, we report on new bounds on the maximal number of abelian gauge group factors in six-dimensional gravitational theories with $\cN=(1,0)$ supersymmetry.
This class of theories is known to be highly constrained by the consistent cancellation of gauge and gravitational anomalies.
Local anomaly considerations have lead to 
a number of remarkable restrictions on the architecture of the gauge sector of six-dimensional supergravities, such as those found in \cite{Kumar:2010ru,Park:2011wv,Taylor:2018khc} (see \cite{Taylor:2011wt} for background and more references).
Apart from being interesting in itself, this line of reasoning can be viewed as a first step towards studying similar questions also in a four-dimensional context.

Abelian gauge symmetries play a distinguished role in the swampland programme: Unlike their non-abelian counterparts, they must necessarily couple to gravity \cite{Lee:2018ihr} in order for their anomalies to be cancelled by a Green-Schwarz mechanism.
It is even more surprising that anomalies alone do not seem to give a universal bound on the number of abelian gauge group factors  even in the highly constrained setup of six-dimensional supergravities with eight supercharges.
In \cite{Park:2011wv} it was observed that the number of abelian gauge group factors can in principle be infinite, as far as the structure of gauge-gravitational anomalies is concerned, at least in absence of charged hypermultiplet matter and as long as the theory contains more than eight tensor multiplets in its spectrum.
For $T\leq 8$ tensor multiplets, on the other hand, \cite{Park:2011wv}  bounds  the maximal number of abelian gauge group factors in purely abelian theories to be \footnote{Ref. \cite{Park:2011wv}  also establishes another bound on the number of $U(1)$ factors in presence of non-abelian gauge sectors, which depends in addition on the number of non-abelian matter representations.} 
\bea \label{Tbound}
N &\leq& (T + 2)(T + \frac{7}{2} + ({T^2 - 51 T + \frac{2225}{4}})^{1/2})\,,   
\eea
{which gives the upper bounds
\be
\{54, 81, 107, 134, 160, 185, 211, 236, 260\} \,,
\ee
for $T=0, 1, \ldots, 8$, respectively,} while for $T=0$ the stronger bound $N \leq 17$ has been found.

As pointed out very recently in \cite{Kim:2019vuc}, extra constraints have to be imposed on a 6d supergravity theory by demanding unitarity of the sector of BPS strings coupling to the tensor multiplets.
In this note we observe that the resulting consistency conditions severely constrain the maximal number of possible abelian gauge group factors, {\it whether or not a non-abelian gauge sector is present}.
In a pure supergravity analysis, bounds can be established modulo certain assumptions on the integrality and generic non-degeneracy of a class of BPS strings. With these assumptions, we will bound, in section \ref{sec_SUGRA}, the number of abelian gauge group factors  to be $N \leq 20$ for $T \geq 1$ in presence of charged matter (and $N \leq 22$ if some $U(1)$s have no charged matter) and $N \leq 32$ for $T=0$. This rules out the possibility of an infinite number of abelian gauge group factors, {provides a definite bound for $T=0$ even in presence of a non-abelian gauge sector}, and considerably improves the abovementioned bound  (\ref{Tbound}) for $1\leq T \leq 8$. 

While the supergravity analysis relies on certain assumptions on the string charge spectrum, the situation is even clearer once we turn to explicit realizations of 6d $\cN=(1,0)$ supergravities in F-theory in section \ref{sec_Ftheory}.
For all such F-theory models with $T \geq 1$ we are able to constrain the maximal number of abelian group factors to be 
\be \label{N16constr}
N \leq 16
\ee
 by considering a {\it heterotic} string
obtained by wrapping a D3-brane on a distinguished curve on the F-theory compactification space. The appearance of this heterotic string, which exists even in absence of a perturbative heterotic dual, has played a crucial role in the study of swampland and weak gravity conjectures in \cite{Lee:2018urn,Lee:2018spm,Lee:2019tst,Lee:2019xtm}. Our constraint (\ref{N16constr})  heavily relies on the results of \cite{Lee:2018ihr} concerning the form of the so-called height-pairing of rational sections describing abelian gauge groups in F-theory.

Our main message is that abelian gauge groups in six-dimensional F-theory models with $T \geq 1$ can always be embedded in the $E_8 \times E_8$ current algebra on the above-mentioned heterotic string, even if no {\it perturbative} heterotic dual exists. {This is a notable difference to non-abelian gauge groups. Combined with the weaker bound $N \leq 32$, which we will derive for F-theory models with $T=0$}, this predicts a {\it universal} bound on the number of independent non-torsional rational sections  on any elliptically fibered Calabi-Yau 3-fold. To date no such bound has been derived in the mathematics literature.

In section \ref{sec_speculations} we speculate on extensions of our results to F-theory compactifications to four dimensions: As long as the base of the elliptic fibration is the blowup of a rational fibration, we expect the bound $N \leq 16$ to continue to hold for all abelian gauge group factors which are associated with rational sections. We furthermore conjecture that the bound $N \leq 16$ should hold also for six-dimensional F-theory models with $T=0$, which would {even improve the existing conditional bound $N \leq 17$, obtained from anomaly cancellation in absence of non-abelian gauge dynamics}.
Finally, our results on six-dimensional F-theory with $T\geq1$ should constrain also the charge pattern for abelian gauge groups as these must be embeddable into $E_8 \times E_8$.

\section{Background and Review} \label{section_Review}

Consider a 6d $\cN=(1,0)$ supergravity theory with vector multiplets in gauge group $G=\prod_\iota G_\iota$, where $G_\iota$ are simple non-abelian factors, as well as $T$ tensor multiplets.
Each tensor multiplet contains one anti-chiral tensor $B^-_\alpha$, $\alpha=1,\ldots,T$, along with a real scalar. 
In addition the gravity multiplet contains a self-dual tensor $B^+$.
The structure of gauge and gravitational anomalies of this theory is characterized by the anomaly coefficients $b_{G_\iota}$ and $a$, which can be represented as vectors in $\mathbb R^{1,T}$, endowed with an intersection product of signature $(1,T)$.
The gravitational anomaly coefficient $a$ must satisfy the relation $a \cdot a = 9 - T$. {This is solved for instance} by  the vector $ - a = (3,-1,\ldots,-1) \in \mathbb R^{1,T}$, where we are choosing the inner product
to be $\eta_{A B} = {\rm diag}(1,-1,\ldots,-1)$ (for $A \in \{0, \alpha\}$).
The real scalars in the tensor multiplet are assembled in a vector $j \in {\mathbb R}^{1,T}$ subject to the normalization condition $j \cdot j = 1$.
The inverse gauge coupling square of the vector multiplets is controlled by the scalars in the tensor multiplet as
\be
\frac{1}{g^2_{G_\iota}} = j \cdot b_{G_\iota} \,.
\ee

The tensor fields couple to BPS strings whose worldsheet theory is described by a 2d $\cN=(0,4)$ theory.
From the perspective of the supergravity theory alone the BPS strings are characterized by {an integral charge vector $Q$ in a unimodular lattice \cite{Seiberg:2011dr}} of tension
$j \cdot Q$.
As in \cite{Kim:2019vuc} we are interested in strings which do not correspond to instanton strings
in 6d ${\cal N}=(1,0)$ superconformal field theories (SCFTs) once gravity is decoupled or in Little String Theories (LSTs).
An analysis of the anomaly polynomial of these strings reveals that the left- and right-moving central charges of the 2d SCFTs to which the worldsheet theories flow in the infra-red (IR) are given by (see \cite{Kim:2019vuc} and references therein)
\be \label{cLcR}
c_L = 3 Q \cdot Q - 9 Q \cdot a +2, \, \quad c_R = 3 Q \cdot Q - 3 Q \cdot a \,.
\ee
In these expressions, the contribution from the free hypermultiplet describing the center-of-mass motion of the string has already been removed.
Furthermore, (\ref{cLcR}) relies on the identification of the R-symmetry of the 2d worldsheet SCFT in the IR with the factor $SU(2)_R$ in the decomposition 
$SO(4) = SU(2)_R \times SU(2)_l$, where $SO(4)$ is the rotation group acting on the extended directions transverse to the string. 
This is the correct identification for the strings which are not associated with 6d SCFTs or LSTs \cite{Kim:2019vuc}.

The gauge group $G$ of the 6d supergravity as well as the $SU(2)_l$ symmetry act as flavour symmetries on the string, with 't Hooft anomaly coefficients
\be
k_{G_\iota} = Q \cdot b_{G_\iota} \,, \qquad k_l = \frac{1}{2} (Q \cdot Q + Q \cdot a + 2) \,.
\ee 
In the IR 2d SCFT, these take the role of the Kac-Moody level of the worldsheet current realising the respective flavour symmetry (see e.g. \cite{Benini:2013cda} for background).
{Unitarity} of the worldsheet SCFT demands that $c_R$, $k_{G_\iota}$ and $k_l$ are all non-negative and hence \cite{Kim:2019vuc}
\be \label{conditions1}
Q \cdot Q \geq -1 \,,  \quad Q \cdot Q + Q \cdot a \geq -2 \,, \quad  Q \cdot b_{G_\iota} \geq 0 \,.
\ee
The left-moving central charge $c_L$ receives contributions from the 
 left-moving current algebra associated with the flavour groups $G_\iota$ as
 \be \label{cGdef}
c_G = \sum_\iota c_{G_\iota} := \sum_\iota \frac{k_{G_\iota} \,  {\rm dim}(G_\iota)}{k_{G_\iota}  + h^\vee_{G_\iota}} \,.
\ee
Here $h^\vee_{G_\iota}$ is the dual Coxeter numbers of $G_\iota$.
Hence consistency of the worldsheet SCFT  implies furthermore that \cite{Kim:2019vuc}
\be \label{cGconstraint}
\sum_{\iota} c_{G_\iota}   \leq c_L = 3 Q \cdot Q - 9 Q \cdot a +2 \,.
\ee
This has been used in \cite{Kim:2019vuc} to rule out a number of theories with non-abelian gauge groups, which would otherwise satisfy the 6d anomaly cancellation conditions.

If the 6d supergravity theory has a realisation via F-theory compactified on an elliptic Calabi-Yau 3-fold $Y_3$ with projection
\be
\pi: Y_3 \rightarrow B_2 \,,
\ee
then the BPS strings correspond to solitonic strings from D3-branes wrapping a curve $C$ on the K\"ahler surface $B_2$. 
In this case, one identifies the string charge lattice with the cohomology lattice $H^2(B_2,\mathbb Z)$ and the intersection pairing with the cohomological intersection pairing of signature $(1,T)$.
Furthermore, $-a = \bar K_{B_2}$ with $\bar K_{B_2}$ the anti-canonical class of $B_2$.

The condition for the string not to describe a 6d SCFT instanton string is that it wraps a curve of self-intersection not smaller than $-1$.
The worldsheet theory of such strings can then be obtained by reducing the 4d $\cN=4$ SYM theory on a single D3-brane along $C$ with the help of a topological duality twist \cite{Martucci:2014ema}, as detailed in \cite{Haghighat:2015ega,Lawrie:2016axq}. The spectrum of worldsheet fields can be found in Table~\ref{table2dfields}.
{ Suppose the curve class $C$ wrapped by a D3-brane allows for a decomposition $C = \sum_p n_p C_p$, with $C_p$ effective curve classes and $n_p >0$.
Then the 2d SCFT of the string associated with $C$ can split into the a sum of SCFTs associated with $n_p$ D3-branes wrapping $C_p$ \cite{Haghighat:2015ega}.
If the curve $C$ is irreducible, such a split occurs only for specially tuned values of  the moduli of $C$, while for generic moduli the string flows to the SCFT associated with $C$. 
We will be encountering such a situation in many cases.
As long as the non-degenerate string associated with $C$ is itself not an SCFT instanton or LST string, bounds of the form (\ref{cGconstraint}) can be derived from it at generic points in the moduli space even though at special points in the moduli space the SCFT may degenerate to a sum of instanton strings.} 

\renewcommand{\arraystretch}{1.2} 
\begin{table}
  \centering
  \begin{tabular}{|c|c|c|c|c|c|}
    \hline
      \multicolumn{2}{|c|}{Fermions} &
    {Bosons}     & $(0,4)$ &  Multiplicity \cr\hline\hline
     $2\times {\bf (2,1)}_1$  &$+$& $4\times {\bf (1,1)}_0$
    &{Hyper} & $g - 1 + \bar K_{B_2} \cdot C$      \cr
    
    \hline
      $2\times {\bf (1,2)}_1$ & $+$& ${\bf (2,2)}_{0}$ &  Twisted Hyper  & $1$ \cr
           
      \hline
   $2\times {\bf (1,2)}_{-1}$ &${-}$& & Fermi  &$g(C)$\cr
   \hline\hline 
    ${\bf (1,1)}_{-1}$ & $-$& &{\rm half-Fermi} & $8 \bar K_{B_2} \cdot C$    \cr\hline\hline 
  \end{tabular}
  \caption{Massless spectrum of the effective 2d $\cN=(0,4)$ world-sheet theory of the string that arises from a D3-brane wrapping a genus $g$ curve $C \subset B_2$ (not contained in the discriminant). 
The representations are under $SU(2)_R \times SU(2)_l \times SO(1,1)$ and the $+/-$ signs in the second column denote chiralities.  
\label{table2dfields}}
  \end{table}

\section{Bounds on the number of U(1) factors in supergravity}\label{sec_SUGRA}

After this review we now derive bounds on the number of abelian gauge group factors in 6d $\cN=(1,0)$ supergravity theories.
 We will be using the same constraint (\ref{cGconstraint}) as in \cite{Kim:2019vuc}, but applied to abelian gauge groups.

The discussion can be phrased either in the purely field theoretic language or in the context of F-theory.
We begin in this section with a supergravity approach.
This, however, will only lead to bounds barring various assumptions concerning integrality and non-degeneracy of suitable charge vectors, as we will make explicit. By contrast, in explicit string theoretic realizations in F-theory, studied in the next section, no comparable assumptions have to be made, but these follow directly from the geometry of the compactification.
 
To each abelian gauge group factor $U(1)_i$, $i=1, \ldots, N$, of a 6d $\cN=(1,0)$ supergravity we can associate an anomaly coefficient $b_i$ \cite{Park:2011wv}, which is the counterpart of the anomaly coefficient 
$b_{G_\iota}$ for the non-abelian gauge groups. 
Similarly, each abelian gauge group factor $U(1)_i$ induces a $U(1)_i$ current on the BPS string with charge vector $Q$, whose Kac-Moody level is given by
\be 
k_i = Q \cdot b_i \,,
\ee
with the understanding that $k_i = 0$ corresponds to no current.
The $U(1)_i$ current central charge is
\be
c_i:= c_{U(1)_i} = 1    \qquad {\rm for } \quad  k_i \neq 0 \,.
\ee
Suppose now the theory contains a BPS string with charge vector $Q$ subject to (\ref{conditions1}) such that for every $U(1)_i$ gauge group factor in the 6d supergravity theory the level on the string is positive,
\be \label{kinonzero}
k_i  = Q \cdot b_i > 0 \quad    \forall i = 1, \ldots, N \,.
\ee
{Suppose furthermore that the associated string does not generically degenerate into SCFT instantons or LST strings.}
Then 
the total number $N$ of $U(1)_i$ factors in the 6d supergravity theory is constrained to be
\be \label{Nconstraint}
N = \sum_i c_i \leq c_L = 3 Q \cdot Q - 9 Q \cdot a +2 \,.
\ee

As we will discuss, a string charge vector $Q$ satisfying (\ref{kinonzero}) can always be found {as long as for each $U(1)_i$ there exists some hypermultiplet carrying non-zero $U(1)_i$ charge.
Before showing this, note that
 the latter condition is always satisfied in theories with $T \leq 8$ \cite{Park:2011wv}}:
Recall that the anomaly coefficient $b_i$ appears, among other places, in the condition for the cancellation of the quartic $U(1)_i$ anomaly and the mixed $U(1)_i$-gravitational anomaly as
\bea \label{anomaliesU1-1}
b_i \cdot b_i &=& \frac{1}{3} \sum_I M_I (q^{(i)}_I)^4   \,, \\ \label{anomaliesU1-2}
- a \cdot b_i &=&  \frac{1}{6} \sum_IM_I (q^{(i)}_I)^2  \,,
\eea
where $M_I$ denotes the number of hypermultiplets with $U(1)_i$ charge $q_I^{(i)}$. Clearly the first condition implies that $b_i \cdot b_i \geq 0$.
{ If $b_i \cdot b_i =0$, then by (\ref{anomaliesU1-1}) there is no charged matter, and one can furthermore show \cite{Lee:2018ihr} that there can be no kinetic mixing with the remaining abelian gauge factors $U(1)_j$, $j \neq i$.
In this sense the $U(1)_i$ theory is completely trivial. In any case, theories with $b_i \cdot b_i =0$ cannot occur if $T \leq 8$: The reason is that $b_i \cdot b_i = 0$, together with $a \cdot b_i = 0$, which follows from (\ref{anomaliesU1-1}) and (\ref{anomaliesU1-2}), leads to $9-T = a \cdot a \leq 0$.}

 
 \subsection{Supergravities with $T =0$}
 
 Let us now analyze the bound (\ref{Nconstraint}) on the number $N$ of $U(1)_i$ gauge group factors in supergravity.
We begin with the simplest case $T=0$. In this case $-a=3$ and $j=1$, and we 
consider the string with charge 
\be\label{Q0_T=0}
Q_0 = 1
\ee
with $Q_0^2 = 1$, $Q_0 \cdot a = -3$. Here and in the sequel we will always assume that the charge $Q_0$ is properly quantized so that a corresponding string satisfies the Dirac quantization condition;
this is the first of a number of assumptions which can be manifestly verified in the F-theory realizations presented in the next section.
According to the completeness conjecture \cite{Polchinski:2003bq} all strings in the charge lattice are in the physical spectrum of a quantum gravity theory and hence there exists a string with charge~\eqref{Q0_T=0}. 
 
Since $b_i^2 > 0$ and $j \cdot b_i > 0$, we know that $b_i >0$ and hence $Q_0 \cdot b_i > 0$.
It is furthermore clear  that the string cannot degenerate to a product of SCFT or LST strings since there are no such sectors in a theory with no tensor multiplets.
Therefore the number of $U(1)_i$ factors is bounded as
\be \label{T=0bound}
N  \leq c_L = 32  \,,\qquad  T = 0 \,.
\ee
{This bound is to be compared with the conditional bound $N \leq 17$ derived previously in \cite{Park:2011wv} based entirely on the 6d supergravity anomaly conditions. We emphasize that our bound is universal for $T=0$ supergravities and in particular also works in presence of a non-abelian sector.}
 
 \subsection{Supergravities with $T =1$}
 
 Assume next that $T = 1$.
 There exist two choices for the unimodular charge lattice \cite{Seiberg:2011dr},
 \be
 \Gamma_0 = \begin{pmatrix} 0 & 1 \\ 1 & 0 \end{pmatrix} \quad {\rm or} \quad   \Gamma_1 = \begin{pmatrix} 1 & 0 \\ 0 & -1 \end{pmatrix} \,.
 \ee
Let us first discuss $\Gamma_1$: 
One can always choose, without loss of generality, $-a = (3,-1)$ \cite{Kumar:2010ru}. 
 We consider the string with charge 
 \be \label{Q0T=1}
 Q_0 = (1,-1) \,,
 \ee
which satisfies 
 \be \label{Q0Q00}
 Q_0 \cdot Q_0 = 0 \,, \qquad Q_0 \cdot a = -2
 \ee
and therefore fulfills  the first two conditions in (\ref{conditions1}).
{Let us furthermore assume that  the string with charge $Q_0$ does not degenerate (generically) into several strings including instanton strings.
We will explicitly verify this in the F-theory realisation of these models in the next section, but leave it for now as an assumption from the pure supergravity perspective.} 
Let us make the ansatz
\be
b_i = (b_i^0, b_i^1) \,.
\ee
The condition $b_i \cdot b_i > 0$, known to be valid for $T=1$, enforces that
\be \label{options}
b_i^0 > |b_i^1|  \geq 0   \qquad {\rm or} \qquad b_i^0  < - |b_i^1|  \leq  0 \,.
\ee
The second possibility can be excluded by recalling that
$j \cdot b_i$ denotes the diagonal part of the matrix involving the abelian gauge kinetic couplings and must hence be non-negative whenever $j$ lies in the cone of allowed tensor multiplet VEVs.
Let us approach the boundary of this cone (subject to the normalization $j \cdot j =1$) by sending $j \to (1,0)$.
Then demanding that $j \cdot b_i$ remains non-negative rules out the case $b_i^0 < 0$ and thereby the second option in  (\ref{options}).
We thus conclude that 
\be \label{kiconstraintpos}
k_i = Q_0 \cdot b_i = b_i^0 + b_i^1 > 0 \,.
\ee
This means that each $U(1)_i$ couples to the string, and therefore demanding (\ref{Nconstraint}) implies that the number $N$ of $U(1)_i$ factors in the supergravity theory is bounded as
\be \label{20bounda}
N \leq c_L = 20  \,,\qquad  (T = 1) \,.
\ee

For the lattice $\Gamma_0$, there exist two inequivalent choices for $a$ subject to $a \cdot a = 8$, namely
$-a = (2,2)$ or $-a = (4,1)$ \cite{Seiberg:2011dr}.
Either way, $b_i \cdot b_i >0$ implies that in $b_i = (b_i^0,b_i^1)$ both entries must be non-vanishing and hence $Q \cdot b_i \neq 0$ for any $Q = (Q^0, Q^1)$.
If $-a = (2,2)$, one then obtains the same bound (\ref{20bounda}) for $Q_0 = (1,0)$ with $Q_0 \cdot Q_0 = 0$ and $Q_0 \cdot (-a) = 2$ and with the convention for the cone of scalar fields $j$ that $b_i^1 > 0$ to satisfy $j \cdot b_i > 0$.
If $-a = (4,1)$, this choice would imply $Q_0 \cdot Q_0 = 0$ and $Q_0 \cdot (-a) = 1$. If this $Q_0$ is properly quantized, it is clear that this cannot have a geometric interpretation in terms of a curve class on a K\"ahler surface in an F-theory context (for which the Riemannn-Roch theorem would have to hold). Indeed, the choice $-a = (4,1)$ does not have a known string theoretic realisation \cite{Taylor:2018khc}.
In any event, we would obtain the bound $N \leq 11$ from $Q_0$ for $-a = (4,1)$, and by comparison with (\ref{20bounda}), $N \leq 20$ remains to be the relevant constraint for $T=1$ supergravities.

We will give an interpretation of this constraint in the next section.
It is furthermore interesting to note that (\ref{20bounda}) is much stronger than the bound
\be \label{BoundTP}
N \leq (T + 2)(T + \frac{7}{2} + ({T^2 - 51 T + \frac{2225}{4}})^{1/2})\,,
\ee 
which had been found in \cite{Park:2011wv} for a theory with only abelian gauge group factors and $T \leq 8$:
For $T =1$, (\ref{BoundTP}) merely constrains the number of abelian gauge group factors to be $N \leq 81$.

\subsection{Supergravities with $T >1$}

It is clear that even for $1 < T \leq 8$, these results can be generalized,
{where for simplicity we only consider supergravity models with charge lattice 
\be \label{Gammachoice}
\Gamma = {\rm diag}(1,-1,\ldots,-1) \,, \quad -a = (3,-1,\ldots,-1) \,.
\ee 
}
Consider the string with charge 
\be
Q_0 = (1,-1,0,\ldots,0) 
\ee
and make the ansatz $b_i = (b_i^0,b_i^1,b_i^2,\ldots,b_i^T)$.
Again $b_i \cdot b_i > 0$ implies 
(\ref{options}), and the second option can be excluded by demanding $j \cdot b_i >0$ for the boundary value $j = (1,0,\ldots,0)$.
Hence (\ref{kiconstraintpos}) continues to hold, barring the assumption that $Q_0$ is an integer charge vector corresponding to a non-degenerate string.

A similar analysis, again subject to the assumptions concerning integrality and non-degeneracy of charge vectors, can be carried out also for supergravities with $T \geq 9$ and (\ref{Gammachoice}). If $b_i \cdot b_i \neq 0$ for every $U(1)_i$ factor, the same bound~\eqref{20bounda} applies as in the theories with $1 \leq T \leq 8$ and $-a = (3,-1,\ldots,-1)$. Even in the presence of $b_i$'s with $b_i \cdot b_i = 0$, a relaxed bound of $22$ can be obtained by considering $T$ strings with charges
\be
Q_0^{(\alpha)} = (1, 0, \dots, 0, -1, 0, \dots, 0) \,, ~~\text{for}~\alpha=1, \dots, T\,,
\ee
for which the only non-zero entries are the $0^{\rm th}$ and the $\alpha^{\rm th}$ components: Upon combining the resulting constraints, one can bound the total number of $U(1)$ factors by
\bea
N \leq 20 \,\frac{T}{T-1} \leq 22.5 \,,
\eea
where $T \geq 9$ has been used in the last step. In particular, this allows us to rule out models with an infinite number of abelian gauge groups with $b_i \cdot b_i = 0$. As shown in \cite{Park:2011wv}, this possibility cannot be ruled out based solely on the 6d supergravity anomaly cancellation conditions.

{
In summary, for the choices of charge lattice made above, the consistency conditions from BPS strings lead to a considerably stronger bound on the number of abelian gauge group factors
than 6d anomaly considerations alone.
}

\section{Bounds on the number of U(1) factors in F-theory}\label{sec_Ftheory}

We now interpret and sharpen the bounds found in the previous section, focusing on those supergravity theories with a realisation as an F-theory compactification on an elliptic Calabi-Yau 3-fold $\pi: Y_3 \to B_2$.
In order to avoid non-minimal singularities in the fiber, which would spoil the Calabi-Yau condition, the base $B_2$ of the elliptic fibration must be of one of the following three types \cite{Grassi}:
\begin{enumerate}
\item $B_2 = \mathbb P^2 ~\Rightarrow~ T=0$
\item $B_2 = \mathbb F_n$ with $n=0,\ldots,8,12$ 
$~\Rightarrow~ T=1$
\item $B_2 = {\rm Bl}^k(\mathbb F_n)$, a $k$-fold blowup of $\mathbb F_n$ $~\Rightarrow~ T=1+k$
\end{enumerate}
In this note, whenever we speak of F-theory constructions we specifically have one of these three {\it smooth} classes of base spaces in mind.

Abelian gauge potentials $A_i$ in the effective action arise by expanding the M-theory 3-form $C_3$ as
$C_3 = \sum_i A_i \wedge {\tw}_i$, where ${\tw}_i \in H^2(Y_3,\mathbb R)$ is the image of a rational section $S_i$ of $Y_3$ under the so-called Shioda map, ${\tw}_i = \sigma(S_i)$.
The rational sections of $Y_3$ form the Mordell-Weil group MW$(Y_3)$, whose rank therefore counts the number of independent (non-Cartan) $U(1)_i$ gauge factors in the effective action.
See \cite{Weigand:2018rez,Cvetic:2018bni} and references therein for more information.

The anomaly coefficients $b_i$ associated with such $U(1)_i$ gauge factors are given by the 
  so-called height-pairing 
\be
b_i = - \pi_\ast(\sigma(S_i)\cdot \sigma(S_i)) \,.
\ee
The crucial property which we will be using is that, as discussed in \cite{Lee:2018ihr}, based on the considerations in \cite{CoxZucker}, this divisor class can always be written as
\be \label{biexplform}
b_i = \frac{1}{m^2_i} (2 \bar K_{B_2} + \delta_i) \,, \quad \text{for some $m_i \in \mathbb Z_{>0}$}\,,
\ee
where $\delta_i$ is an {\it effective divisor} on $B_2$ and $\bar K_{B_2}$ represents the anti-canonical class on $B_2$.
The effective divisor $b_i$ can be thought of as the linear combination of 7-brane divisors  supporting the abelian gauge group $U(1)_i$.

We can now interpret and sharpen the constraints found in the previous section for the three types of F-theory base spaces listed above.

\subsection{F-theory Vacua with $T=0$}
For $T=0$, i.e. $B_2 = \mathbb P^2$, we can consider a string by wrapping a D3-brane along the rational curve $C = H$, where $H$ denotes the hyperplane class of $B_2$ with $H \cdot H=1$, in terms of which $\bar K_{\mathbb P^2} = 3 H$. 
Every height-pairing $b_i$ can be parametrized as $b_i = b H$ for $b>0$. 
This system is of course an explicit realisation of the supergravity setup with $T=0$ considered in the previous section and 
simply reproduces the bound (\ref{T=0bound}). We will come back to this setup in section \ref{sec_speculations} and argue that this bound can be sharpened.

\subsection{F-theory Vacua with $T=1$}

For $T=1$, the F-theory base is a Hirzebruch surface $B_2= \mathbb F_n$, for $n=0,1,\ldots,8,12$, corresponding to a fibration of a rational curve  $\mathbb P^1_f$  over a base $\mathbb P^1_h$ \footnote{For $n=9,10,11$ additional blowups are required \cite{Morrison:2012np}.}.
The classes of the fiber, $f$, and of the base, $h$, span the cohomology $H^2(\mathbb F_n,\mathbb Z) = \langle f, h \rangle$ and have the intersection numbers
\be
f \cdot f = 0 \,,\qquad f \cdot h = 1 \,, \qquad h \cdot h = -n \,.
\ee
 The Mori cone of effective curves, ${\bf M}(\mathbb F_n)$, and the closure of the K\"ahler cone, ${\bf \bar K}(\mathbb F_n)$, are given by 
 \be \label{MFcones}
{\bf M}(\mathbb F_n) = \langle f , h \rangle \,,\quad 
{\bf \bar K}(\mathbb F_n) = \langle f , h + n \, f \rangle
\ee 
while the anti-canonical class evaluates to
\be
{\bar K}_{\mathbb F_n} = 2h + (2 + n) f \,.
\ee

Consider now the string obtained by wrapping a D3-brane along  the curve class
\be
C = f 
\ee
with
\be \label{CdotCproperties}
C \cdot C = 0 \,, \qquad C \cdot \bar K_{\mathbb F_n} = 2 \,.
\ee
This is an explicit realisation of the properties (\ref{Q0Q00}) for an integral curve class $C$.
 Since $C=f$ is a {\it generator} of the Mori cone, it cannot split into other effective curve classes. The worldsheet SCFT associated with the string is therefore manifestly non-degenerate. 
Since $f$ is furthermore in the closure of the K\"ahler cone (\ref{MFcones}), we know that 
\be
f \cdot \delta \geq 0    \qquad \forall \delta \quad  {\rm effective} \,.
\ee
As a result, any $U(1)_i$ factor in the theory leads to a worldsheet current on the string with a non-trivial Kac-Moody level because
\be \label{kicomptuation}
k_i = C \cdot b_i = f \cdot \frac{1}{m_i^2} (2 \bar K_{\mathbb F_n} + \delta_i) \geq \frac{2}{m_i^2} f \cdot \bar K_{\mathbb F_n} = \frac{4}{m_i^2} > 0 \,,
\ee
where we used (\ref{biexplform}).
Hence the string associated with $C$ satisfies all properties to apply (\ref{Nconstraint}) to bound the number $N$ of abelian gauge group factors in such F-theory models, leading again to the bound
\be 
N \leq c_L = 20 \,.
\ee

In fact, the string from the D3-brane wrapped on $C=f$ is not just any solitonic string: 
Its zero-mode spectrum coincides with the spectrum of a critical heterotic string propagating in six dimensions; see Table~\ref{table2dfields} for the field contents of the world-sheet theory. 
This is of course no surprise because F-theory on $B_2 = \mathbb F_n$ is dual to the heterotic string on an elliptic $K3$ over $\mathbb P^1_h$, and in this duality the solitonic string from a D3-brane wrapping the fiber $\mathbb P^1_f$ reduces to the critical heterotic string in the limit of small volume of $\mathbb P^1_f$. This realisation of the heterotic string has played an important role in the context of the weak gravity conjecture in F-theory \cite{Lee:2018urn,Lee:2018spm,Lee:2019tst, Lee:2019xtm}.

The value $c_L = 20$ of the heterotic string has an immediate interpretation: 
The left-moving sector of the string (apart from the free hypermultiplet associated to the string motion in $\mathbb R^{1,5}$) comprises $8 C \cdot \bar K_{{\mathbb F}_n} =16$ left-moving fermions in $\cN=(0,4)$ half-Fermi multiplets, along with one hypermultiplet associated with the propagation of the string on the dual heterotic K3 surface (see Table \ref{table2dfields}). 
The half-Fermi multiplets are due to the zero modes at the intersection of $C$ with the 7-branes on $B_2$. They are the only source of charge of the string with respect to the 7-brane gauge groups since the hypermultiplet is uncharged.
This suggests that the number of $U(1)_i$ gauge fields is in fact subject to the slightly stronger bound
\be \label{16bounda}
N \leq (c_L - 4) = 16  \,,
\ee
where we have subtracted the contribution from the ($U(1)_i$ neutral) hypermultiplet associated with the moduli of the string along the dual K3 surface. 
In hindsight, the bound  (\ref{16bounda}) is  simply the well-known statement that the $U(1)_i$ gauge groups in models with a {\it perturbative} heterotic dual {\it on a smooth K3} can be embedded into the perturbative heterotic $E_8 \times E_8$ 
gauge group. A far less trivial statement is that the bound $N \leq 16$ continues to hold in all F-theory compactifications with $T > 1$, as we will see momentarily.

It is interesting to compare the bound (\ref{16bounda}) relevant for geometric F-theory vauca to the less strict bound (\ref{20bounda}) derived for general supergravity theories with $T=1$.
Consistent supergravities satisfying (\ref{20bounda}) but not (\ref{16bounda}) exist.
For instance, 6d ${\cal N}=(1,0)$ vacua of the perturbative heterotic string  can have a gauge group of rank up to $20$ as is known from purely conformal field theoretic constructions (e.g. \cite{Lerche:1986cx}). These are, however, not the {\it geometric} heterotic K3 compactifications which the solitonic string in our F-theory model is related to \footnote{We thank W. Lerche for discussions on this point.}. Correspondingly, a split of the worldsheet fields of a heterotic string as in Table \ref{table2dfields}, which underlies the reduced bound (\ref{16bounda}), 
is a consequence of the specific realisation of the heterotic worldsheet theory as a wrapped D3-brane in F-theory setups.
In this sense, the bound $N \leq 20$ found in the supergravity approach, without any prejudice concerning an embedding into F-theory, is indeed the correct bound for general 6d ${\cal N}=(1,0)$ supergravities.

Finally, recall the well-known fact that F-theory on $\mathbb F_n$ necessarily supports a non-Higgsable non-abelian gauge group $G_n$ on 7-branes wrapping the base class $h$ with self-intersection $h \cdot h = -n$.
Since $b_{G_n} \cdot C = h \cdot f = 1$, this gives rise to a $G_n$ current on the string at level $k_{G_n} =1$, which contributes to the central charge $c_L$ with $c_{G_n}$ given in (\ref{cGdef}). 
From Table~\ref{table_Fn}, we find that this improves the bound on the number of possible abelian gauge group factors to be
\be \label{boundimproved}
N \leq \lfloor 16 - c_{G_n}\rfloor = \{16,16,16,14,12,10,10,9,9,8 \} \,,
\ee
for F-theory models on $B_2 =\mathbb F_n$ with $n = 0,1,\ldots,8,12$, respectively.

\begin{table}
  \centering
  \begin{tabular}{|c|||c|c|c|c|c|c|c|c|c|c|}
    \hline
$\mathbb F_n$                     & 0 & 1 & 2 & 3 & 4 & 5 & 6 & 7 & 8 & 12 \\ \hline
$G_n$                 & $-$ & $-$ & $-$ & $SU(3)$ & $SO(8)$ & $F_4$ & $E_6$ & $E_7$ & $E_7$ & $E_8$ \\ \hline
$h_G^\vee$ & $-$ & $-$ & $-$ & $3$ & $6$ & $9$ & $12$ & $18$ & $18$ & $30$ \\ \hline
 \end{tabular}
  \caption{Non-Higgsable gauge groups on $\mathbb F_n$.}  \label{table_Fn}
  \end{table}

\subsection{F-theory Vacua with $T>1$}

We now come to the most interesting case,  where the F-theory base $B_2$ is an arbitrary blowup of $\mathbb F_n$.
Recall that  $\mathbb F_n$ is a non-degenerate $\mathbb P^1_f$ fibration.
After we blow up a single point into an exceptional divisor (curve), the new fiber degenerates into two curves when it hits the blowup locus.
This process can be repeated multiple times.
Let us introduce the morphism
\be \label{bmap}
p: B_2 \to \mathbb F_n \,,
\ee
which corresponds to the blowdown of the exceptional divisors $E_l$ on $B_2$.
The anti-canonical classes of both spaces are related as $ \bar K_{B_2} = p^\ast(\bar K_{{\mathbb F}_n}) + \sum_l a_l E_l$ for some coefficients $a_l$.
The idea is now to consider the string obtained from a D3-brane wrapping the curve
\be
C = p^\ast(f) \subset B_2 \,,
\ee
where $f$ is the fiber class on $\mathbb F_n$ prior to the blowup.
This curve has the properties
\be
C \cdot_{B_2} C = p^\ast(f) \cdot_{B_2} p^\ast(f) = f \cdot_{\mathbb F_n} f = 0
\ee
and 
\bea
C \cdot_{B_2} \bar K_{B_2}  &=&   p^\ast(f) \cdot_{B_2}(p^\ast(\bar K_{{\mathbb F}_n}) + \sum_l a_l E_l) \\
&=&    f \cdot_{{\mathbb F_n}} \bar K_{{\mathbb F}_n} +  \sum_l a_l f \cdot_{\mathbb F_n} p_\ast(E_l)   \\
&=& f \cdot_{{\mathbb F_n}} \bar K_{{\mathbb F}_n}  = 2 \,.
\eea
Before the last line we have used that $ f \cdot_{\mathbb F_n} p_\ast(E_l) = 0$ because the blow-down map $p$ contracts the exceptional divisors to points.

The string associated with $C$ therefore still has the same zero-mode structure as the heterotic string in six dimensions.
As alluded to above, the difference to the situation with $T=1$ is that the curve $C=p^{*}(f)$ may split, at certain loci in its moduli space, into various exceptional divisors on $B_2$. 
In the dual heterotic picture, this corresponds to a degeneration of the heterotic string into various non-critical strings at the location of NS5-branes.
Importantly for us, away from the NS5-branes, the heterotic worldsheet is non-degenerate and continues to be described by a single SCFT in the IR for which the analysis of section \ref{section_Review} is valid. 

It is still true that each $U(1)_i$ gauge symmetry in the 6d F-theory induces an abelian worldsheet current on the string of non-vanishing level $k_i \neq 0$.
This is because prior to the blowup, $f$ is in the closure ${\bf \bar K}(\mathbb F_n)$ of the K\"ahler cone, and its pullback under the blowdown (\ref{bmap}) continues to lie in the K\"ahler cone closure~\cite{GH}. 
Therefore, the same computation as in (\ref{kicomptuation}), now performed on $B_2$, shows that 
\be \label{kibigger0}
k_i = C \cdot_{B_2} b_i > 0 \qquad \forall \, b_i \,.
\ee
We conclude that the same upper bound $N \leq c_L = 20$, or in fact
\be \label{finalbound}
N \leq (c_L-4) = 16
\ee
holds for {\it any} F-theory compactification with at least one tensor (or in fact the stronger bound (\ref{boundimproved}) for F-theory on a $k$-fold blowup of $\mathbb F_n$). 
Here we have again subtracted the contribution from the $U(1)_i$ uncharged hypermultiplet on the worldsheet of the heterotic string as realized by a wrapped D3-brane.

Note that this does certainly not mean that the total rank of the gauge group, including non-abelian factors, should be bounded by $16$: Unlike abelian gauge group factors, 
non-abelian gauge symmetries can `hide' from the worldsheet of the heterotic string in that they need not necessarily induce a worldsheet current with non-zero level $k_{G_\iota}$.
After all, the crucial property of abelian gauge symmetries underlying our analysis is that their anomaly coefficients take the form (\ref{biexplform}), which prevents $k_i = 0$ for the heterotic string under consideration.
Non-abelian gauge group factors with $k_{G_\iota} = 0$ with respect to the heterotic string are truly non-perturbative in that they cannot be embedded into the heterotic $E_8 \times E_8$ current on the heterotic worldsheet. 
The anomaly coefficients of such gauge groups satisfy 
\be
b_{G_\iota} \cdot b_{G_\iota} \leq 0 
\ee
because $b_{G_\iota} \cdot C = 0$ and $C \cdot C = 0$ can only be solved, with an intersection form of signature $(1,T)$, for $b_{G_\iota} \cdot b_{G_\iota} \leq 0$ (and $b_{G_\iota} \cdot b_{G_\iota} = 0$ if and only if $b_{G_\iota} = \alpha  C$ for some $\alpha \in \mathbb Q$). 
In fact, $b_{G_\iota}$ is the divisor class wrapped by the 7-brane with gauge group $G_\iota$, and the gauge groups with $b_{G_\iota} \cdot C = 0$ include all the gauge groups wrapping blowup-up divisors on $B_2$: As discussed, these blowup divisors are contained in $C=p^{*}(f)$ and hence intersect it trivially. These exceptional divisors support 6d SCFT sectors on shrinkable curves on $B_2$ and are, in the above sense, truly non-perturbative from the heterotic string point of view~\footnote{This is to be contrasted with the 6d SCFT sectors on the base $\mathbb P^1_h$ of $\mathbb F_n$ which do induce a worldsheet current on the heterotic string.}.
In particular, our results imply that it is not possible to higgs these to abelian gauge group factors. This is not only (trivially) true  for the non-Higgsable gauge groups supported on blowup divisors, but also after enhancing the gauge group further by tunings of the elliptic fibration in F-theory. Even after such tunings enhancing $G \to H$, there are no adjoint scalars available to higgs $H \to G \times U(1)^l$ because the blowup curves are rigid.

Another way to phrase the origin of the bound (\ref{finalbound}) is this:
F-theory on a blow-up of $\mathbb F_n$ is dual to the heterotic string on K3 with NS5-branes. Due to non-perturbative effects associated with the NS5-branes, additional gauge groups will in general occur which cannot be 
embedded into $E_8 \times E_8$. In F-theory these are realized on the blow-up divisors, and since the latter have vanishing intersection with the curve $C = p^*(f)$, they do not correspond to a current algebra on the heterotic string.
On the other hand, abelian gauge groups are guaranteed to induce a $U(1)_i$ current on the worldsheet because of (\ref{kibigger0}). Hence the non-perturbative sector does not contain the $U(1)_i$ gauge groups, which are consequently purely perturbative and embeddable into $E_8 \times E_8$.

\section{Discussion and Speculations} \label{sec_speculations}

In this short note, we have derived universal bounds on the number $N$ of abelian gauge group factors in F-theory compactifications to six dimensions.
Our main result is the bound $N \leq 16$ for any 6d F-theory compactification giving rise to at least one tensor multiplet.
The derivation combines the recent idea of \cite{Kim:2019vuc} to deduce constraints on the form of 6d $\cN=(1,0)$ supergravity theories from consistency of embedded BPS strings on the one hand, and specific properties of the anomaly coefficients for abelian gauge group factors \cite{Lee:2018ihr,CoxZucker} on the other hand. 
As in \cite{Lee:2018urn,Lee:2018spm,Lee:2019tst,Lee:2019xtm} we have identified a certain solitonic string on any F-theory base other than $\mathbb P^2$ as the heterotic string;
applying the constraint (\ref{cGconstraint}) to this string in the context of abelian gauge groups leads to the precise bounds (\ref{boundimproved})  and (\ref{finalbound}). 
For F-theory on an elliptic fibration over base $\mathbb P^2$, no such heterotic string can be identified in the spectrum of solitonic strings; the resulting bound found for this class of models hence a priori only takes the form $N \leq 32$. We will come back to this point below.

Our results are interesting from the perspective of the swampland programme of ruling out theories without a UV completion despite their (apparent) low-energy consistency:
For example, in \cite{Park:2011wv} it had been observed that 6d anomaly cancellation alone is not enough to rule out 6d supergravity models with an infinite number of $U(1)$ gauge fields with anomaly coefficients $b_i \cdot b_i = 0$.
Our analysis shows that such theories are in the swampland, both from the perspective of F-theory and, barring the assumptions stated, more generally in supergravity even without invoking any string theoretic  realisation. 

Our results can also be interpreted as giving a prediction for the maximal rank of the Mordell-Weil group of any elliptically fibered Calabi-Yau threefold to be
\be \label{rkMWbound}
{\rm rk}({\rm MW}(Y_3)) \leq 16 \, \, (B_2 \neq \mathbb P^2)   \quad {\rm or}  \quad \leq 32 \, \, (B_2 = \mathbb P^2) \,,
\ee 
where the bound for $B_2 = \mathbb P^2$ will be revisited below.
To date, the elliptically fibered Calabi-Yau threefold with the highest known Mordell-Weil rank is an elliptic fibration over ${\mathbb P}^2$ with ${\rm rk}({\rm MW}(Y_3))  = 10$ \cite{Elkies}. It would be interesting to see if the bounds (\ref{rkMWbound}) for elliptic Calabi-Yau 3-folds can be improved further, or else be derived from a purely mathematical point of view.

We end with a number of extensions of our results, some of which are speculative. \vspace{1pt} \\

\noindent {\bf F-theory on $\mathbb P^2$ Revisited}

First, it is natural to conjecture that the universal bound $N \leq 16$ should also apply to F-theory with base $\mathbb P_2$.
The total value of $c_L = 32$ for the string associated with $C= H$ on $\mathbb P^2$ can be understood as $c_L = 24 + 8$, where the second factor is due to the two interacting hypermultiplets of the string (see again Table \ref{table2dfields}) and the first factor is due to the $8 \bar K_{\mathbb P^2} \cdot C = 24$ fermionic $3-7$ modes.
Since these are the only source of $U(1)_i$ charge of the string, it is again suggestive that $N \leq c_L - 8 = 24$. 

However, we conjecture that this is still an over-counting.
To understand its origin in our framework, note that the 
surface $\mathbb P^2$ can be obtained from $\mathbb F_1$ by blowing down the base curve $\mathbb P^1_h$ with class $h$.
On $\mathbb F_1$, we can consider the string associated with the curve class $C' = f + h$.
By blowing down $h$, $C'$ maps to the class $C=H$ associated with the string on $\mathbb P^2$ which we are interested in.
 By wrapping a D3-brane along $C'$ on $\mathbb F_1$ we obtain a bound state of the heterotic string (from a D3-brane on $f$) with one E-string (from a D3-brane on $h$ with $h \cdot h = -1$). 
At generic values of the curve moduli the string is non-degenerate and described by the invariants
$C'\cdot C' = 1$, $C' \cdot \bar K_{\mathbb F_1} =3$. This matches the string on the curve $C=H$ after the blowdown on $\mathbb P^2$.
The total left-moving central charge for the string on $C'$ is  $c_L = 32 = 24 + 8$. 
Since $C' = h + f$ is in the K\"ahler cone (\ref{MFcones}), we know that 
 $k_i=b_i \cdot C' \neq 0$ for every $U(1)_i$ in the theory, but $b_i \cdot h$ need not be non-vanishing. 
 Indeed, the 
 heterotic string has a current algebra $E^{(1)}_8 \times E^{(2)}_8$, while the $E$-string has only a single $E_8$ current, and some $U(1)_i$ factors are orthogonal to this $E_8$.
The bound $N \leq 24$ can then be interpreted as $N \leq N_{f} + N_h$, where $N_f=16$ is the number of $U(1)_i$ which couple to the heterotic string and $N_h$ is the number of $U(1)_i$
coupling to the E-string. Since the latter set is included in the first, we end up with $N \leq 16$ for the total number of $U(1)_i$.
Now, on $\mathbb P^2$, there exists no individual heterotic and $E$-string, but only their bound state along $C=H$; nonetheless we conjecture that a similar overcounting is at work, and that the bound can be improved to $N \leq 16$, which is even sharper than {the conditional bound $N \leq 17$ from anomalies~\cite{Park:2011wv}}.
This translates into the conjecture that ${\rm rk}({\rm MW}(Y_3)) \leq 16$ for any elliptic Calabi-Yau 3-fold with base $\mathbb P^2$, and 
it would be extremely interesting to verify or falsify this speculation. \vspace{1pt} \\

\noindent {\bf Extensions to 4d F-theory}

Our findings can be immediately extended to F-theory compactifications on Calabi-Yau 4-folds to four dimensions. This is particularly useful because no comparable anomaly constraints exist
in this case, at least for (non-chiral) situations without gauge fluxes.

Strings from D3-branes wrapping curves $C$ on the base three-fold $B_3$ are now described by a 2d ${\cal N}=(0,2)$ worldsheet theory \cite{Lawrie:2016axq}, even though these strings are not BPS objects from the viewpoint of the 
4d ${\cal N}=1$ supersymmetry algebra. As long as $B_3$ is the blowup of a $\mathbb P^1_f$-fibration, a D3-brane wrapping the class of the generic $\mathbb P^1_f$ fiber gives rise to a heterotic string, which becomes the fundamental string in the dual heterotic frame \cite{Lee:2019tst}. This continues to hold even if no {\it perturbative} heterotic dual of the F-theory model exists due to the presence of NS5-branes on the heterotic side. These now wrap curves on the base of the dual heterotic  Calabi-Yau 3-fold, which are the location over which the blowups have been performed on the F-theory side.
The form (\ref{biexplform}) for the $U(1)_i$ anomaly coefficients $b_i = \frac{1}{m_i^2}(2 \bar K_{B_3} + \delta_i)$ carries over to Calabi-Yau 4-folds and by similar arguments as before, every $U(1)_i$ couples to the heterotic worldsheet in F-theory.
More precisely, prior to the blowup, let us denote by $r: B_3 \to B_2$ the $\mathbb P^1_f$ fibration of $B_3$, with $B_2$ a surface; then we can always realize the fiber $\mathbb P^1_f$ as the class $\mathbb P^1_f = \frac{1}{n}(r^\ast J_0 \cdot r^\ast J_0)$ for some 
K\"ahler form $J_0$ on $B_2$ (and $n$ an integer). Now, $r^\ast J_0 \cdot r^\ast J_0 \cdot \delta_i \geq 0$ because $J_0$ pulls back to an element in the closure of the K\"ahler cone of $B_3$ and $\delta_i$ is effective, and 
$r^\ast J_0 \cdot r^\ast J_0 \cdot \bar K_{B_3} =2$ because the fiber is a rational curve of self-intersection zero.
Hence $k_i = b_i \cdot [\mathbb P^1_f] \geq  4/m_i^2 > 0 $. This result remains invariant under  blowups of $B_3$.

We therefore reproduce the same bound 
\be \label{Nbound4d}
N \leq c_L  - 6 = 22-6 =16 \,,
\ee
where $c_L=22$ denotes the left-moving central charge of the internal sector (after subtracting the contribution from $2$ free scalars corresponding to the motion in the two transverse directions) and  the subtraction of $6$ corrects for the chiral left-moving scalars which are uncharged under the 7-brane gauge group, as before.  See Table 3.1. of \cite{Lee:2019tst} for details of the worldsheet spectrum.

Note that (\ref{Nbound4d}) only constrains the number of abelian gauge group factors which are realized in terms of rational sections, on blowups of rational fibrations.
Unlike in six dimensions, even the part of the non-abelian gauge sector which is not embedded in the heterotic worldsheet algebra may contribute additional $U(1)$s  due to the possibility of breaking the gauge group with gauge fluxes. 
These sectors are not constrained by our arguments. Such U(1)s, however, are not realized by rational sections of the fibration.
Furthermore the set of possible base spaces is considerably richer for four-dimensional F-theory, and it would be interesting to extend our bounds to other geometries. 
\vspace{1pt} \\

\noindent {\bf Constraints on Charges }

Coming back to six dimensions for simplicity, the essence of our bounds is the realisation that abelian gauge groups are embedded into the $E_8 \times E_8$ algebra of a heterotic worldsheet, at least for models with $T \geq 1$.
This statement holds independently of the existence of a perturbative heterotic dual.
It should
 also carry over to the level of charged matter states in that not only 
 the possible number of abelian gauge group factors is constrained, but also the possible spectrum of charges. 
In other words, all abelian gauge charges in F-theory models with $T \geq 1$ should follow from a decomposition of the ${\bf 248} \times  {\bf 248}$ of $E_8 \times E_8$. 
It would be very interesting to understand this further, in particular in view of recent attempts of classifying possible charges in F-theory geometry (e.g. in \cite{Baume:2015wia,Lawrie:2015hia,Cvetic:2015ioa,Taylor:2018khc,Raghuram:2018hjn}, with a more complete list of references provided in \cite{Weigand:2018rez,Cvetic:2018bni}). \vspace{2mm}\\


{\noindent \bf Acknowledgements}
We thank Wolfgang Lerche for crucial discussions and for collaboration on related topics. 
The work of SJL is supported by the Korean Research Foundation (KRF) through the CERN-Korea Fellowship program.

\end{document}